\documentclass[%
 reprint,
superscriptaddress,
nofootinbib,
bibnotes,
 amsmath,amssymb,
 aps,
]{revtex4-1}

\usepackage{graphicx}
\usepackage{dcolumn}
\usepackage{braket}
\usepackage{gensymb}
\usepackage{bm}
\usepackage{amsmath}
\usepackage{float}
\usepackage{mathtools}
\usepackage{color}
\usepackage{csquotes}
\usepackage[utf8]{inputenc}
\usepackage{thmtools}

\declaretheorem[]{theorem}

\newcommand{\ps}{{\rm ps}}

\usepackage{hyperref}

\usepackage[normalem]{ulem}

\makeatletter
\def\p@subsection{}
\makeatother
\makeatletter
\def\p@subsubsection{}
\makeatother

\allowdisplaybreaks

\begin{document}

\preprint{APS/123-QED}

\title{Quantum Advantage in Postselected Metrology}
\author{David R. M. Arvidsson-Shukur}
\affiliation{ Cavendish Laboratory, Department of Physics, University of Cambridge, Cambridge CB3 0HE, United Kingdom
}
\affiliation{%
Department of Mechanical Engineering, Massachusetts Institute of Technology, Cambridge, Massachusetts 02139, USA}
\affiliation{%
Research Laboratory of Electronics, Massachusetts Institute of Technology, Cambridge, Massachusetts 02139, USA
}%

\author{Nicole Yunger Halpern}
\affiliation{ITAMP, Harvard-Smithsonian Center for Astrophysics, Cambridge, MA 02138, USA}
\affiliation{Department of Physics, Harvard University, Cambridge, MA 02138, USA}
\affiliation{%
Research Laboratory of Electronics, Massachusetts Institute of Technology, Cambridge, Massachusetts 02139, USA
}%

\author{Hugo V. Lepage}
\affiliation{%
 Cavendish Laboratory, Department of Physics, University of Cambridge, Cambridge CB3 0HE, United Kingdom
}%
\author{Aleksander A. Lasek}
\affiliation{%
 Cavendish Laboratory, Department of Physics, University of Cambridge, Cambridge CB3 0HE, United Kingdom
}%
\author{Crispin H. W. Barnes}
\affiliation{%
 Cavendish Laboratory, Department of Physics, University of Cambridge, Cambridge CB3 0HE, United Kingdom
}%
\author{Seth Lloyd}
\affiliation{%
Department of Mechanical Engineering, Massachusetts Institute of Technology, Cambridge, Massachusetts 02139, USA}
\affiliation{%
Research Laboratory of Electronics, Massachusetts Institute of Technology, Cambridge, Massachusetts 02139, USA
}%

\date{\today}

\begin{abstract}
We show that postselection offers a nonclassical advantage in metrology. In every parameter-estimation experiment, the final measurement or the postprocessing incurs some cost. Postselection can improve the rate of Fisher information (the average information learned about an unknown parameter from an experimental trial) to cost. This improvement, we show, stems from the negativity of a quasiprobability distribution, a quantum extension of a probability distribution. In a classical theory, in which all observables commute, our quasiprobability distribution can be expressed as real and nonnegative. In a quantum-mechanically noncommuting theory, nonclassicality manifests in negative or nonreal quasiprobabilities. The distribution's nonclassically negative values enable postselected experiments to outperform even postselection-free experiments whose input states and final measurements are optimized: Postselected quantum experiments can yield anomalously large information-cost rates. We prove that this advantage is genuinely nonclassical: no classically commuting theory can describe any quantum experiment that delivers an anomalously large Fisher information. Finally, we outline a preparation-and-postselection procedure that can yield an \textit{arbitrarily} large Fisher information. Our results establish the nonclassicality of a metrological advantage, leveraging our quasiprobability distribution as a mathematical tool.
\end{abstract}

\maketitle


\section*{Introduction}

Our ability to deliver new quantum-mechanical improvements to technologies relies on a better understanding of the foundation of quantum theory: When is a phenomenon truly nonclassical? \color{black} We take noncommutation as our notion of nonclassicality and we quantify this nonclassicality with  negativity: \color{black} Quantum states can be represented by quasiprobability distributions, extensions of classical probability distributions. Whereas probabilities are real and nonnegative, quasiprobabilities can assume negative and nonreal values. Quasiprobabilities' negativity stems from the impossibility of representing quantum states with joint probability distributions \cite{Lutkenhaus95, Spekkens08, Ferrie08}. The distribution we use, an extension of the Kirkwood-Dirac distribution \cite{Kirkwood33, Dirac45, Yunger18}, signals nonclassical noncommutation through the presence of negative or nonreal quasiprobabilities.

 One field advanced by quantum mechanics is metrology, which concerns the statistical estimation of unknown physical parameters. Quantum metrology relies on quantum phenomena to improve estimations beyond classical bounds \cite{Giovanetti11}. A famous example exploits entanglement \cite{Giovanetti06, Krischek11, Demkowicz14}. Consider using $N$ separable and distinguishable probe states to evaluate identical systems in parallel. The best estimator's error will scale as $N^{-1/2}$. If the probes are entangled, the error scaling improves to $N^{-1}$ \cite{Maccone13}. As Bell's theorem rules out classical (local realist) explanations of entanglement, the improvement is genuinely quantum. 

A central quantity in parameter estimation is the Fisher information, $\mathcal{I}(\theta)$. The Fisher information quantifies the average information learned about an unknown parameter $\theta$ from an experiment \cite{Helstrom76, Braunstein94, bCover06}. $\mathcal{I}(\theta)$  lower-bounds the variance of an unbiased estimator $\theta_e$ via the Cramér-Rao inequality: $\textrm{Var}( \theta_e ) \geq 1 / \mathcal{I}(\theta)$ \cite{Cramer16, Rao92}. A common metrological task concerns optimally estimating a parameter that characterizes a physical process. The experimental input and the final measurement are optimized to maximize the Fisher information and to minimize the estimator’s error.

\begin{figure}
\includegraphics[scale=0.60]{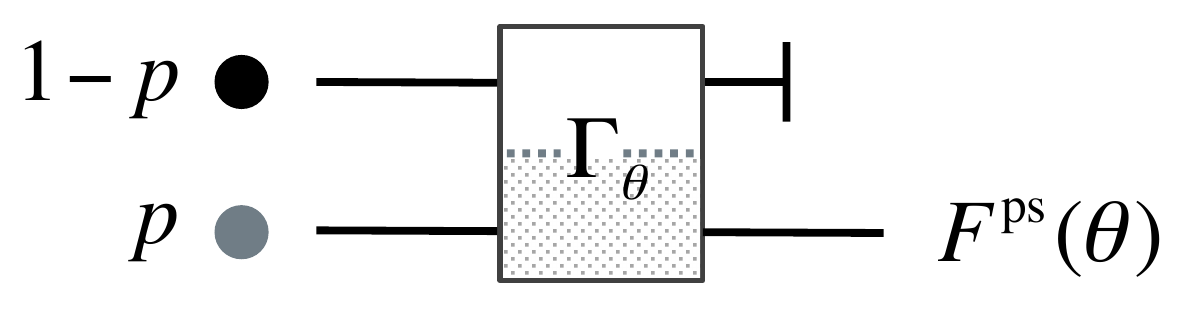}
\caption{\textbf{Classical experiment with postselection.} A nonoptimal input device initializes a particle in one of two states, with probabilities $p$ and $1 - p$, respectively. The particle undergoes a transformation $\Gamma_{\theta}$ set by an unknown parameter $\theta$. Only the part of the transformation that acts on particles in the lower path depends on $\theta$. If the final measurement is expensive, the particles in the upper path should be discarded: they possess no information about $\theta$. }
\label{fig:ClassicalPS}
\end{figure}

Classical parameter estimation can benefit from postselecting the output data before postprocessing. Postselection can raise the Fisher information per final measurement or postprocessing event (Fig. \ref{fig:ClassicalPS}). Postselection can also raise the rate of information per final measurement in a quantum setting. But classical postselection is intuitive, whereas an intense discussion surrounds postselected quantum experiments \cite{Vaidman88, Leifer05, Tollaksen07, Aharonov08, Dressel10, Vaidman13, Ferrie14, Pusey14, Pusey15, ArvShukur16, ArvShukur17-2, Schmid18, ArvShukur19, Cimini20}. The ontological nature of postselected quantum states, and the extent to which they exhibit nonclassical behavior, is subject to an ongoing debate. Particular interest has been aimed at pre- and postselected averages of observables. These \textit{weak values} can lie outside an observable's eigenspectrum when measured via a weak coupling to a pointer particle \cite{Vaidman88, Duck89}. Such values offer metrological advantages in estimations of weak-coupling strengths \cite{Tollaksen07, Dressel14, Pang14, Pusey14, Jordan14, Harris17, Kunjwal18, Xu20}.

In this article, we go beyond this restrictive setting and ask, can postselection provide a nonclassical advantage in \textit{general} quantum parameter-estimation experiments? We conclude that it can. We study metrology experiments for estimating an unknown transformation parameter whose final measurement or postprocessing incurs an experimental cost \cite{Liuzzo18, Lipka18}. Postselection allows the experiment to incur that cost only when the postselected measurement's result reveals that the final measurement's Fisher information will be sufficiently large. We express the Fisher information in terms of a quasiprobability distribution.  Quantum negativity in this distribution enables postselection to increase the Fisher information above the values available from standard input-and-measurement-optimized experiments. Such an anomalous Fisher information can improve the rate of information gain to experimental cost, offering a genuine quantum advantage in metrology. We show that, within a commuting theory, a theory in which observables commute classically, postselection can improve information-cost rates no more than a strategy that uses an optimal input and final measurement can. We thus conclude that experiments that generate anomalous Fisher-information values require noncommutativity.

\section*{Results}

\subsection*{Postselected quantum Fisher information} \label{Sec:FishInf}

As aforementioned, postselection can raise the Fisher information per final measurement. Figure \ref{fig:ClassicalPS} outlines a classical experiment with such an information enhancement. Below, we show how postselection affects the Fisher information in a quantum setting.

Consider an experiment with outcomes $ i $ and associated probabilities $p_i(\theta)$, which depend on some unknown parameter $\theta$. The Fisher information about $\theta$ is \cite{bCover06} 
\begin{equation}
\label{Eq:ClFish}
\mathcal{I}(\theta) = \sum_{i} p_i(\theta) [\partial_{\theta} \ln(p_i(\theta))]^2 = \sum_{i} \frac{1}{p_i(\theta)} [\partial_{\theta} p_i(\theta)]^2 .
\end{equation}
Repeating the experiment $N \gg 1$ times provides, on average, an amount $N \mathcal{I}(\theta)$ of information about $\theta$. The estimator's variance is bounded by $\textrm{Var}( \theta_e ) \geq 1 / [N \mathcal{I}(\theta)]$.

Below, we define and compare two types of metrological procedures. In both scenarios, we wish to estimate an unknown parameter $\theta$ that governs a physical transformation.

\textbf{Optimized prepare-measure experiment:} An input system undergoes the partially unknown transformation, after which the system is measured. Both the input system and the measurement are chosen to provide the largest possible Fisher information.

\textbf{Postselected prepare-measure experiment:} An input system undergoes, first, the partially unknown transformation and, second, a postselection measurement. Conditioned on the postselection's yielding the desired outcome, the system undergoes an information-optimized final measurement.

In quantum parameter estimation, a quantum state is measured to reveal information about an unknown parameter encoded in the state. We now compare, in this quantum setting, the Fisher-information values generated from the two metrological procedures described above. Consider a quantum experiment that outputs a state $\hat{\rho}_{\theta} = \hat{U}(\theta) \hat{\rho}_0 \hat{U}^{\dagger}(\theta)$, where $\hat{\rho}_0 $ is the input state and $\hat{U}(\theta)$ represents a unitary evolution set by $\theta$. The quantum Fisher information is defined as the Fisher information maximized over all possible generalized measurements \cite{Braunstein94, Fujiwara95, bPetz07, bPetz11, Giovanetti11}:
\begin{equation}
\mathcal{I}_Q(\theta | \hat{\rho}_{\theta}) = \mathrm{Tr} \big[ \hat{\rho}_{\theta} \hat{\Lambda}_{\hat{\rho}_{\theta}}^2 \big] .
\end{equation}
$\hat{\Lambda}_{\hat{\rho}_{\theta}}$ is the symmetric logarithmic derivative, implicitly defined by $\partial_{\theta} \hat{\rho}_{\theta} = \frac{1}{2}(\hat{\Lambda}_{\hat{\rho}_{\theta}} \hat{\rho}_{\theta}+\hat{\rho}_{\theta} \hat{\Lambda}_{\hat{\rho}_{\theta}})$ \cite{Helstrom76}. 

If $\hat{\rho}_{\theta}$ is pure, such that $\hat{\rho}_{\theta} =\ket{\Psi_{\theta}} \bra{\Psi_{\theta}}$, the quantum Fisher information can be written as \cite{Pang14, Pang15}
\begin{align}
\label{Eq:FishQ}
\mathcal{I}_Q(\theta| \hat{\rho}_{\theta}) =  4 \braket{\dot{\Psi}_{\theta}|\dot{\Psi}_{\theta}} - 4 |\braket{\dot{\Psi}_{\theta}|\Psi_{\theta}}|^2 ,
\end{align}
where $ \ket{\dot{ \Psi}_{\theta}}  \equiv \partial_{\theta}  \ket{\Psi_{\theta}} $.

We assume that the evolution can be represented in accordance with Stone's theorem \cite{Stone32}, by $\hat{U}(\theta) \equiv e^{-i \hat{A} \theta}$, where $\hat{A}$ is a Hermitian operator. We assume that $\hat{A}$ is not totally degenerate: If all the $\hat{A}$ eigenvalues were identical, $\hat{U}(\theta)$ would not imprint $\theta$ onto the state in a relative phase. For a pure state, the quantum Fisher information equals $\mathcal{I}_Q(\theta | \hat{\rho}_{\theta}) = 4 \textrm{Var}(\hat{A})_{\hat{\rho}_{0}}$ \cite{Giovanetti11}. Maximizing Eq. \ref{Eq:ClFish} over all measurements gives $\mathcal{I}_Q(\theta | \hat{\rho}_{\theta})$. Similarly, $\mathcal{I}_Q(\theta | \hat{\rho}_{\theta})$ can be maximized over all input states. For a given unitary $\hat{U}(\theta) = e^{-i \hat{A} \theta}$, the maximum quantum Fisher information is 
\begin{equation}
\label{Eq:MaxFish}
\mathrm{max}_{\hat{\rho}_0} \big\{ \mathcal{I}_Q(\theta | \hat{\rho}_{\theta}) \big\} = 4 \mathrm{max}_{\hat{\rho}_0} \big\{ \textrm{Var}(\hat{A})_{\hat{\rho}_{0}} \big\} =  (\Delta a)^2 ,
\end{equation}
where $\Delta a$ is the difference between the maximum and minimum eigenvalues of $\hat{A}$ \cite{Giovanetti11}.\footnote{  The information-optimal input state is a pure state in an equal superposition of one eigenvector associated with the smallest eigenvalue and one associated with the largest. } To summarize, in an optimized quantum prepare-measure experiment, the quantum Fisher information is $(\Delta a)^2$.

\begin{figure}
\includegraphics[scale=0.45]{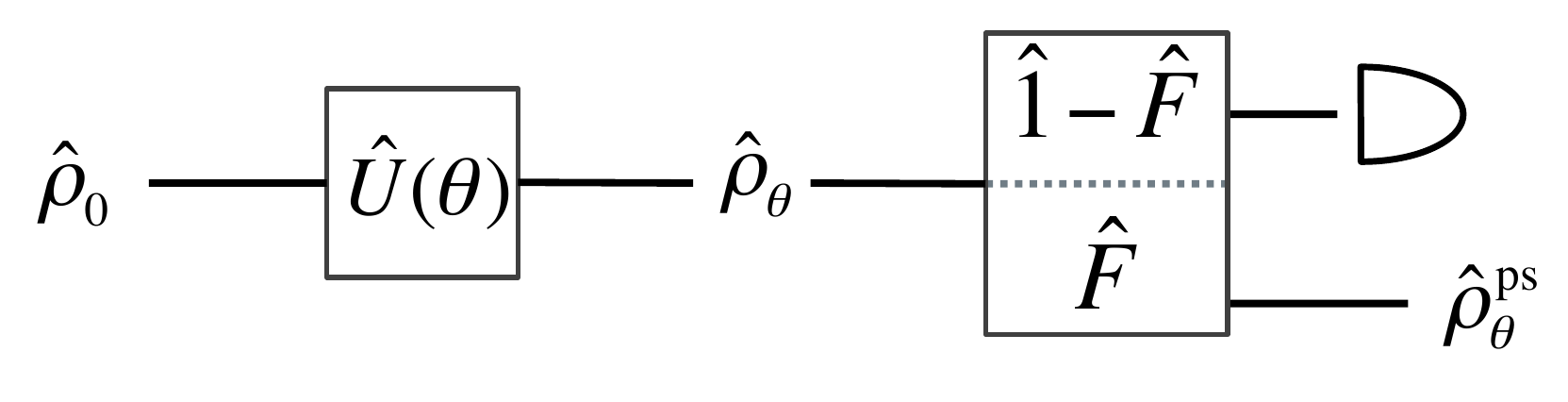}
\caption{\textbf{Preparation of postselected quantum state.} \color{black} First, an input quantum state $\hat{\rho}_0$ undergoes a unitary transformation $\hat{U}(\theta) = e^{-i \theta \hat{A}}$: $\hat{\rho}_0 \rightarrow \hat{\rho}_{\theta}$. Second, the quantum state is subject to a projective postselective measurement $\{\hat{F}, \hat{1} - \hat{F} \}$. The postselection is such that if the outcome related to the operator $\hat{F}$ happens, then the quantum state is not destroyed. The experiment outputs renormalized states $\hat{\rho}_{\theta}^{\textrm{ps}} = \hat{F}\hat{\rho}_{\theta} \hat{F} / \mathrm{Tr}(\hat{F}\hat{\rho}_{\theta})$. \color{black}}
\label{fig:EncodingPS}
\end{figure}

We now find an expression for the quantum Fisher information in a postselected prepare-measure experiment. A projective postselection occurs after $\hat{U}(\theta)$ but before the final measurement. Figure \ref{fig:EncodingPS} shows such a quantum circuit. The renormalized quantum state that passes the postselection is $\ket{\Psi_{\theta}^\ps} \equiv   \ket{\psi_{\theta}^\ps} / \sqrt{p_{\theta}^{\ps}}$, where we have defined an unnormalized state $ \ket{\psi_{\theta}^\ps}  \equiv \hat{F} \ket{\Psi_{\theta}} $ and the postselection probability $p_{\theta}^\ps \equiv \mathrm{Tr}(\hat{F}\hat{\rho}_{\theta})$. As before, $\hat{\rho}_{\theta} = \hat{U}(\theta) \hat{\rho}_0 \hat{U}^{\dagger}(\theta)$.  $\hat{F} = \sum_{f \in \mathcal{F}^\ps} \ket{f} \bra{f}$ is the postselecting projection operator, and $\mathcal{F}^\ps$ is a set of orthonormal basis states allowed by the postselection.  Finally, the postselected state undergoes an information-optimal measurement.

When $\ket{\Psi_{\theta}^\ps} \equiv   \ket{\psi_{\theta}^\ps} / \sqrt{p_{\theta}^{\ps}}$ is substituted into Eq. \ref{Eq:FishQ}, the derivatives of $p_{\theta}^\ps$ cancel, such that
\begin{equation}
\label{Eq:FishQpsShort}
\mathcal{I}_Q(\theta|\Psi_{\theta}^\ps) =   4\braket{\dot{\psi}_{\theta}^\ps|\dot{\psi}_{\theta}^\ps} \frac{1}{p_{\theta}^\ps} -  4 |\braket{\dot{\psi}_{\theta}^\ps|\psi_{\theta}^\ps}|^2 \frac{1}{(p_{\theta}^\ps)^2} .
\end{equation}
Equation \ref{Eq:FishQpsShort} gives the quantum Fisher information available from a quantum state after its postselection. 
Unsurprisingly, $\mathcal{I}_Q(\theta|\Psi_{\theta}^\ps)$ can exceed $\mathcal{I}_Q(\theta|\hat{\rho}_{\theta})$, since $p_{\theta}^\ps \leq 1$. Also classical systems can achieve such postselected information amplification (see Fig. \ref{fig:ClassicalPS}). Unlike in the classical case, however, $\mathcal{I}_Q(\theta|\Psi_{\theta}^\ps)$ can also exceed the Fisher information of an optimized prepare-measure experiment, $(\Delta a)^2$. We show how below.

\subsection*{Quasiprobability representation} \label{Sec:Quasi}

In classical mechanics, our knowledge of a point particle can be described by a probability distribution for the particle's position, $\vec{x}$, and momentum, $\vec{k}$: $p(\vec{x}, \vec{k})$. In quantum mechanics, position and momentum do not commute, and a state cannot generally be represented by a joint probability distribution over observables' eigenvalues. A quantum state can, however, be represented by a quasiprobability distribution. Many classes of quasiprobability distributions exist. The most famous is the Wigner function \cite{Wigner32, Wootters87, Carmichael13}. Such a distribution satisfies some, but not all, of Kolmogorov's axioms for probability distributions \cite{Kolmogorov51}: the entries sum to unity, and marginalizing over the eigenvalues of every observable except one yields a probability distribution over the remaining observable's eigenvalues. A quasiprobability distribution can, however, have negative or nonreal values. Such values signal nonclassical physics in, for example, quantum computing and quantum chaos \cite{Spekkens08, Ferrie11, Kofman12, Howard14, Dressel15, Delfosse15, Delfosse17, Yunger17, Yunger18, Yunger18-2, Yunger19}.

A cousin of the Wigner function is the Kirkwood-Dirac quasiprobability distribution \cite{Kirkwood33, Dirac45, Yunger18}. \color{black} This distribution, which has been referred to by several names across the literature, resembles the Wigner function for continuous systems. Unlike the Wigner functions, however, the Kirkwood-Dirac distribution is well-defined for discrete systems, even qubits.  The Kirkwood-Dirac distribution has been used in the study of weak-value amplification \cite{Steinberg95, Johansen07, Hofmann12-2, Dressel15, Piacentini16, Yunger18}, information scrambling \cite{Yunger18, Yunger18-2, Yunger19, Razieh19} and direct measurements of quantum wavefunctions \cite{Lundeen11, Lundeen12, Bamber14, Thekkadath16}. Moreover, negative and nonreal values of the distribution have been linked to nonclassical phenomena \cite{Dressel15, Yunger18, Yunger18-2, Yunger19}. \color{black} We cast the quantum Fisher information for a postselected prepare-measure experiment in terms of a \textit{doubly extended}\footnote{The modifier ``doubly extended'' comes from the experiment in which one would measure the distribution: One would prepare $\hat{\rho}$, sequentially measure two observables weakly, and measure one observable strongly. The number of weak measurements equals the degree of the extension \cite{Yunger18}.} Kirkwood-Dirac quasiprobability distribution \cite{Yunger18}.  \color{black} We employ this distribution due to its usefulness as a mathematical tool: This distribution enables the proof that, in the presence of noncommuting observables, postselection \color{black} can give \color{black} a metrological protocol a nonclassical advantage.

Our distribution is defined in terms of eigenbases of $\hat{A}$ and $\hat{F}$. Other quasiprobability distributions are defined in terms of bases independent of the experiment. For example, the Wigner function is often defined in the bases of the quadrature of the electric field or the position and momentum bases. However, basis-independent distributions can be problematic in the hunt for nonclassicality \cite{Spekkens08, Delfosse15}. Careful application, here, of the extended Kirkwood-Dirac distribution ties its nonclassical values to the operational specifics of the experiment.
\color{black} 

To begin, we define the quasiprobability distribution of an arbitrary quantum state $\hat{\rho}$:
\begin{equation}
\label{Eq:DefQuas}
q_{a,a^{\prime},f}^{\hat{\rho}} \equiv \braket{f|a}\bra{a}\hat{\rho} \ket{a^{\prime}} \braket{a^{\prime} | f} .
\end{equation}
Here, $\{ \ket{f} \}$,  $\{ \ket{a} \}$ and $\{ \ket{a^{\prime}} \}$ are bases for the Hilbert space on which $\hat{\rho}$ is defined.  We can expand $\hat{\rho}$ \cite{Lundeen11, Lundeen12} as\footnote{If any $\braket{f|a} = 0$, we perturb one of the bases infinitesimally while preserving its orthonormality.}
\begin{align}
\label{Eq:DefQuasRho}
\hat{\rho}
& =  \sum_{a,a^{\prime},f} \frac{\ket{a}\bra{f}}{\braket{f|a}} q_{a,a^{\prime},f}^{\hat{\rho}}  .
\end{align}  

Let $\{ \ket{a} \} = \{ \ket{a^{\prime}} \}$ denote an eigenbasis of $\hat{A}$, and let $\{ \ket{f} \}$ denote an eigenbasis of $\hat{F}$.  \color{black} The reason for introducing a doubly extended distribution, instead of the standard Kirkwood-Dirac distribution $q_{a,f}^{\hat{\rho}} \equiv \braket{f|a}\bra{a}\hat{\rho} \ket{ f}$, is  that  $\mathcal{I}_Q(\theta | \Psi_{\theta}^\ps)$ can be expressed most concisely, naturally, and physically meaningfully in terms of $q_{a,a^{\prime},f}^{\hat{\rho}_{\theta}}$. Later, we shall see how the nonclassical entries in $q_{a,a^{\prime},f}^{\hat{\rho}}$ and $q_{a,f}^{\hat{\rho}}$ are related.   \color{black} We \color{black} now \color{black} express the postselected quantum Fisher information (Eq. \ref{Eq:FishQpsShort}) in terms of the quasiprobability values $q_{a,a^{\prime},f}^{\hat{\rho}_{\theta}}$ (Supp. Inf. $1$).
\begin{align}
\label{Eq:FishQuas}
\mathcal{I}_Q(\theta | \Psi_{\theta}^\ps) =  4  \sum_{
\substack{a,a^{\prime}, \\ f\in \mathcal{F}^\ps } 
} \frac{q_{a,a^{\prime},f}^{\hat{\rho}_{\theta}}}{p_{\theta}^{\ps}}   a a^{\prime}   - 4   \Big| \sum_{
\substack{a,a^{\prime}, \\ f\in \mathcal{F}^\ps } 
} \frac{q_{a,a^{\prime},f}^{\hat{\rho}_{\theta}}}{p_{\theta}^{\ps}}  a \Big|^2 ,
\end{align}
where $a$ and $a^{\prime}$ denote the eigenvalues associated with $\ket{a}$ and $\ket{a^{\prime}}$, respectively.\footnote{We have suppressed degeneracy parameters $\gamma$ in our notation for the states, e.g., $\ket{a, \gamma}  \equiv  \ket{a}$.} Equation \ref{Eq:FishQuas} contains a conditional quasiprobability distribution, $q_{a,a^{\prime},f}^{\hat{\rho}_{\theta}} / p_{\theta}^{\ps}$.
If $\hat{A}$ commutes with $\hat{F}$, as they do classically, then they share an eigenbasis for which $q_{a,a^{\prime},f}^{\hat{\rho}_{\theta}} / p_{\theta}^{\ps} \in [0, \, 1] $, and the postselected quantum Fisher information is bounded as $ \mathcal{I}_Q(\theta | \Psi^\ps_{\theta}) \leq (\Delta a)^2$: 
\begin{theorem}
In a classically commuting theory, no postselected prepare-measure experiment can generate more Fisher information than the optimized prepare-measure experiment.
\end{theorem}

\textit{Proof of Theorem 1.---}We upper-bound the right-hand side of Eq. \ref{Eq:FishQuas}. First, if $ \{ \ket{a} \} = \{ \ket{a^{\prime}} \} = \{ \ket{f} \}$ is a eigenbasis shared by $\hat{A}$ and $\hat{F}$, Eq. \ref{Eq:DefQuas} simplifies to a probability distribution:
\begin{equation}
q_{a,a^{\prime},f}^{\hat{\rho}_{\theta}} = \bra{a} \hat{\rho}_{\theta} \ket{a^{\prime}} \bm{[}\ket{f} = \ket{a} \bm{]} \bm{[}\ket{a^{\prime}} = \ket{f} \bm{]}  \in [0, \, 1] ,
\end{equation} where $\bm{[} X \bm{]}$ is the Iverson bracket, which equals $1$ if $X$ is true and equals $0$ otherwise. Second, summing $q_{a,a^{\prime},f}^{\hat{\rho}_{\theta}} / p_{\theta}^{\ps}$ over $f\in \mathcal{F}^\ps$, we find  
\begin{equation}
\sum_{f\in \mathcal{F}^\ps} q_{a,a^{\prime},f}^{\hat{\rho}_{\theta}} / p_{\theta}^{\ps}  = \bra{a} \hat{\rho}_{\theta} \ket{a^{\prime}} \bra{a^{\prime} } \hat{F} \ket{a} / p_{\theta}^{\ps}.
\end{equation} By the eigenbasis shared by $\hat{A}$ and $\hat{F}$, the sum simplifies to $\bra{a} \hat{\rho}_{\theta} \hat{F} \ket{a^{\prime}} \bm{[}\ket{a^{\prime}} = \ket{a} \bm{]}   / p_{\theta}^{\ps}$. We can thus rewrite Eq. \ref{Eq:FishQuas}: 
\begin{align}
\mathcal{I}_Q(\theta | \Psi_{\theta}^\ps) = & 
 4  \sum_{
\substack{a,a^{\prime} } 
} \frac{ \bra{a} \hat{\rho}_{\theta} \hat{F} \ket{a^{\prime}} \bm{[}\ket{a^{\prime}} = \ket{a} \bm{]}   }{p_{\theta}^{\ps}}   a a^{\prime}  \nonumber \\ & - 4   \Big| \sum_{
\substack{a,a^{\prime}} 
} \frac{ \bra{a} \hat{\rho}_{\theta} \hat{F} \ket{a^{\prime}} \bm{[}\ket{a^{\prime}} = \ket{a} \bm{]}   }{p_{\theta}^{\ps}}  a \Big|^2
  \nonumber \\
  = &
4   \sum_{a} q_a a^2    - 4   \Big( \sum_{a}q_a  a \Big)^2 , \label{Eq:NonNegFish1}
\end{align}
where we have defined the probabilities $q_a \equiv \bra{a} \hat{\rho}_{\theta} \hat{F} \ket{a}    / p_{\theta}^{\ps} = \sum_{f\in \mathcal{F}^\ps} \bra{a} \hat{\rho}_{\theta}  \ket{a} \bm{[}\ket{f} = \ket{a} \bm{]}     / p_{\theta}^{\ps}$. 

Apart from the multiplicative factor of $4$, Eq. \ref{Eq:NonNegFish1} is in the form of a variance with respect to the observable's eigenvalues $ a $. Thus, Eq. \ref{Eq:NonNegFish1} is maximized when $q_{a_{\textrm{min}}} = q_{a_{\textrm{max}}} = \frac{1}{2}$:
\begin{align}
\max_{\{ q_a\}} \{ \mathcal{I}_Q(\theta | \Psi_{\theta}^\ps) \}  = (\Delta a)^2 .
\end{align}
This Fisher-information bound must be independent of our choice of eigenbases of $\hat{A}$ and $\hat{F}$.  In summary, if $\hat{A}$ commutes with $\hat{F}$, then all $q_{a,a^{\prime},f}^{\hat{\rho}_{\theta}} / p_{\theta}^{\ps}$ can be expressed as real and nonnegative, and $\mathcal{I}_Q(\theta | \Psi_{\theta}^\ps) \leq (\Delta a)^2 $. $\square$


In contrast, if the quasiprobability distribution contains negative values, the postselected quantum Fisher information can violate the bound: $\mathcal{I}_Q(\theta | \Psi^\ps_{\theta}) > ( \Delta a)^2$. 
In Supp. Inf. $2$, we prove a second theorem:\footnote{The theorem's converse is not generally true.}
 \begin{theorem}
An anomalous postselected Fisher information implies that the quantum Fisher information cannot \color{black} decompose \color{black} in terms of a  nonnegative doubly extended Kirkwood-Dirac quasiprobability distribution.
\end{theorem}
\textit{Proof:} see Supplementary Note 2 for a proof.

\color{black}
This inability to \color{black} decompose \color{black} implies that $\hat{A}$ fails to commute with $\hat{F}$. However, pairwise noncommutation of $\hat{\rho}_{\theta}$, $\hat{A}$ and $\hat{F}$ is insufficient to enable anomalous values of $\mathcal{I}_Q(\theta | \Psi^\ps_{\theta}) $. For example, noncommutation could lead to \color{black} a  nonreal Kirkwood-Dirac distribution without any negative real components. Such a distribution \color{black}  cannot improve $\mathcal{I}_Q(\theta | \Psi^\ps_{\theta}) $ beyond classical values.  Furthermore, the presence or absence of commutation is a binary measure. In contrast, how much postselection improves $\mathcal{I}_Q(\theta | \Psi^\ps_{\theta}) $ depends on how much negativity $q_{a,a^{\prime},f}^{\hat{\rho}_{\theta}} / p_{\theta}^{\ps}$ has. We build on this observation, and  propose two experiments that yield anomalous Fisher-information values, in Supp. Infs. $3$ and $4$.\color{black}\footnote{ \color{black} It remains an open question to investigate the relationship between Kirkwood-Dirac negativity in other metrology protocols with noncommuting operators, e.g., \cite{Sun20}. \color{black}}

\color{black} As promised, we now address the relation between nonclassical entries in $q_{a,a^{\prime},f}^{\hat{\rho}}$ and nonclassical entries in $q_{a,f}^{\hat{\rho}}$. \color{black} For pure states $\hat{\rho} = \ket{\Psi} \bra{\Psi}$, the doubly extended quasiprobability distribution can be \color{black} expressed time symmetrically \color{black} in terms of the standard Kirkwood-Dirac distribution \cite{Kirkwood33, Dirac45, Yunger17, Yunger18, Yunger18-2, Yunger19}: $q_{a,a^{\prime},f}^{\hat{\rho}} = \frac{1}{p_{f}} q_{a,f}^{\hat{\rho}} \big( q_{a^{\prime},f}^{\hat{\rho}} \big)^{*}$,
where  $q_{a,f}^{\hat{\rho}} = \braket{f|a}\bra{a}\hat{\rho} \ket{ f}$ and $p_{f} \equiv |\braket{f|\Psi}|^2$.\footnote{See \cite{Aharonov08, Leifer17} for discussions about time-symmetric interpretations of quantum mechanics. }  \color{black} Therefore, \color{black} a negative  $q_{a,a^{\prime},f}^{\hat{\rho}}$  implies negative or nonreal values of $q_{a,f}^{\hat{\rho}}$. Similarly, a negative  $q_{a,a^{\prime},f}^{\hat{\rho}}$ implies a negative or nonreal weak value $\braket{f | a} \braket{a | \Psi} / \braket{ f | \Psi } $ \cite{Vaidman88}, which possesses interesting ontological features (see below). Thus, an anomalous Fisher information is closely related to a negative or nonreal weak value. Had we weakly measured the observable $\ket{a}\bra{a}$ of $\hat{\rho}_{\theta}$ with a qubit or Gaussian pointer particle before the postselection, and had we used a fine-grained postselection $\{ \hat{1} - \hat{F}, \, \ket{f}\bra{f} \; : \; f \in \mathcal{F}^\ps \}$, the weak measurement would have yielded a weak value outside the eigenspectrum of $\ket{a}\bra{a}$. It has been shown that such an anomalous weak value proves that quantum mechanics, unlike classical mechanics, is contextual: quantum outcome probabilities can depend on more than a unique set of underlying physical states \cite{Spekkens05, Pusey14, Kunjwal18}. If $\hat{\rho}_{\theta}$ had undergone the aforementioned weak measurement, instead of the postselected prepare-measure experiment, the weak measurement's result would have signaled quantum contextuality. Consequently, a counterfactual connects an anomalous Fisher information and quantum contextuality. While counterfactuals create no problems in classical physics, they can lead to logical paradoxes in quantum mechanics \cite{Kochen75, Hardy92, bPenrose94, Spekkens05}. Hence our counterfactual's implication for the ontological relation between an anomalous Fisher information and contextuality offers an opportunity for future investigation.

\subsection*{Improved metrology via postselection} \label{Sec:Metro}

In every real experiment, the preparation and final measurement have costs, which we denote $\mathcal{C}_{P}$ and $\mathcal{C}_{M}$, respectively. For example, a particle-number detector's \textit{dead time}, the time needed to reset after a detection, associates a temporal cost with measurements \cite{Greganti18}. \color{black} Reference \cite{Liuzzo18} concerns a two-level atom in a noisy environment. Liuzzo \textit{et al.} detail the tradeoff between frequency estimation's time and energy costs. Standard quantum-metrology techniques, they show, do not necessarily improve metrology, if the experiment's energy is capped. \color{black}  Also, the cost of postprocessing can be incorporated into $\mathcal{C}_{M}$.\footnote{In an experiment, these costs can be multivariate functions that reflect the resources and constraints. Such a function could combine a detector's dead time with the monetary cost of liquid helium and a graduate student's salary. However, presenting the costs in a general form benefits this platform-independent work.} We define the information-cost rate as $R(\theta) \coloneqq N \mathcal{I}(\theta) / (N \mathcal{C}_{P} + N \mathcal{C}_{M}) = \mathcal{I}(\theta) / ( \mathcal{C}_{P} +  \mathcal{C}_{M}) $. If our experiment conditions the execution of the final measurement on successful postselection of a fraction $p_{\theta}^{\ps}$ of the states, we include a cost of postselection, $\mathcal{C}_\ps$. We define the postselected experiment's information-cost rate as $R^\ps(\theta) \coloneqq N p_{\theta}^{\ps} \mathcal{I}^\ps(\theta) / (N \mathcal{C}_{P} + N \mathcal{C}_\ps + N p_{\theta}^{\ps}\mathcal{C}_{M}) = p_{\theta}^{\ps} \mathcal{I}^\ps(\theta) / ( \mathcal{C}_{P} + \mathcal{C}_\ps + p_{\theta}^{\ps} \mathcal{C}_{M}) $, where $\mathcal{I}^\ps(\theta)$ is the Fisher information conditioned on successful postselection. Generalizing the following arguments to preparation and measurement costs that differ between the postselected and nonpostselected experiments is straightforward.


In classical experiments, postselection can improve the information-cost rate. See Fig. \ref{fig:ClassicalPS} for an example. But can postselection improve the information-cost rate in a classical experiment with information-optimized inputs? Theorem 1 answered this question in the negative. $ \mathcal{I}^\ps(\theta)  \leq \mathrm{max} \{  \mathcal{I}(\theta) \}$ in every classical experiment. The maximization is over all physically accessible inputs and final measurements. A direct implication is that $  R^\ps(\theta) \leq \mathrm{max} \{ R(\theta) \}$.

In quantum mechanics, $\mathcal{I}_Q(\theta | \Psi^\ps_{\theta})  $ can exceed  $\max_{\hat{\rho}_{0}} \{ \mathcal{I}_Q(\theta | \hat{\rho}_{\theta}) \} = (\Delta a)^2 $. This result would be impossible classically. Anomalous Fisher-information values require quantum negativity in the doubly extended Kirkwood-Dirac distribution. Consequently, even compared to quantum experiments with optimized input states, postselection can raise information-cost rates beyond classically possible rates: $  R^\ps(\theta) > \mathrm{max} \{ R(\theta) \}$. This result generalizes  the metrological advantages observed in the measurements of weak couplings, which also require noncommuting operators. References \cite{Hosten08, Dixon09, Starling09, Brunner10, Starling10, Egan12, Hofmann12, Magana14, Lyons14, Martinez17} concern metrology that involves weak measurements of the following form. The primary system $\textrm{S}$ and the pointer $\textrm{P}$ begin in a pure product state $  \ket{\Psi_{\rm S}}  \otimes \ket{ \Psi_{\rm P} }$; the coupling Hamiltonian is a product $\hat{H}= \hat{A}_{\rm S} \otimes \hat{A}_{\rm P}$; the unknown coupling strength $\theta$ is small; and just the system is postselected. Our results govern arbitrary input states, arbitrary Hamiltonians (that satisfy Stone's theorem), arbitrarily large coupling strengths $\theta$, and arbitrary projective postselections.  Our result shows that postselection can improve quantum parameter estimation in experiments where the final measurement's cost outweighs the combined costs of state preparation and postselection: $\mathcal{C}_{M} \gg \mathcal{C}_{P} + \mathcal{C}_\ps$. \color{black}  Earlier works identified that the Fisher information from nonrenormalized trials that succeed in the postselection cannot exceed the Fisher information averaged over all trials, including the trials in which the postselection fails \cite{Ferrie14-2, Combes14}.\footnote{\color{black} Reference \cite{Pang15} considered  squeezed coherent states  as metrological  probes in specific  weak-measurement experiments.  It is shown that postselection can improve the signal-to-noise ratio, irrespectively of whether the analysis includes the failed trials. However, this work concerned nonpostselected experiments in which only the probe state was measured. Had it been possible to successfully measure also the target system, the advantage would have disappeared. \color{black}} \color{black} In accordance with practical metrology, not only the Fisher information, but also measurements' experimental costs, underlie our results. 

So far, we have shown that $\mathcal{I}_Q(\theta | \Psi^\ps_{\theta})$ can exceed $( \Delta a)^2$. But how large can $\mathcal{I}_Q(\theta | \Psi^\ps_{\theta})$ grow?  In Supp. Inf. $3$, we show that, if the generator $\hat{A}$ has $M \geq 3$ not-all-identical eigenvalues, there is no upper bound on $\mathcal{I}_Q(\theta | \Psi^\ps_{\theta})$. If  $ \mathcal{C}_{P}$ and $ \mathcal{C}_\ps$ are negligible compared to $\mathcal{C}_{M}$, then there is no theoretical cap on how large $R^\ps(\theta) $ can grow. In general, when $\mathcal{I}_Q(\theta | \Psi^\ps_{\theta}) \rightarrow \infty$,  $p_{\theta}^{\ps} \times \mathcal{I}_Q(\theta | \Psi^\ps_{\theta}) < ( \Delta a)^2$, such that information is lost in the events discarded by postselection. But if $\hat{A}$ has doubly degenerate minimum and maximum eigenvalues, $p_{\theta}^{\ps} \times \mathcal{I}_Q(\theta | \Psi^\ps_{\theta})$ can approach $( \Delta a)^2$ while $\mathcal{I}_Q(\theta | \Psi^\ps_{\theta})$ approaches infinity (see Supp. Inf. $4$). In such a scenario, postselection can improve information-cost rates, as long as $\mathcal{C}_\ps < ( 1 - p_{\theta}^{\ps} ) \mathcal{C}_{M} $---a significantly weaker requirement than $\mathcal{C}_{M} \gg \mathcal{C}_{P} + \mathcal{C}_\ps$.

\section*{Discussion} \label{Sec:Disc}

From a practical perspective, our results highlight an important quantum asset for parameter-estimation experiments with expensive final measurements. In some scenarios, the postselection's costs exceed the final measurement's costs, as an unsuccessful postselection might require fast feedforward to block the final measurement. But in single-particle experiments, the postselection can be virtually free and, indeed, unavoidable: an unsuccessful postselection can destroy the particle, precluding the triggering of the final measurement's detection apparatus \cite{Calafell19}. Thus, current single-particle metrology could benefit from postselected improvements of the Fisher information. A photonic experimental test of our results is currently under investigation.

From a fundamental perspective, our results highlight the strangeness of quantum mechanics as a noncommuting theory. Classically, an increase of the Fisher information via postselection can be understood as the \textit{a posteriori} selection of a better input distribution. But it is nonintuitive that quantum mechanical postselection can enable a quantum state to carry more Fisher information than the best possible input state could. The optimized Cramér-Rao bound, obtained from Eq. \ref{Eq:MaxFish}, can be written in the form of an uncertainty relation: $\sqrt{\textrm{Var}(\theta_e)} (\Delta a) \geq 1$ \cite{Giovanetti11}. Our results highlight the probabilistic possibility of violating this bound. More generally, the information-cost rate's ability to violate a classical bound leverages negativity, a nonclassical resource in quantum foundations, for metrological advantage.

\subsection*{Acknowledgements}
The authors would like to thank Justin Dressel, Nicolas Delfosse, Matthew Pusey, Noah Lupu-Gladstein, Aharon Brodutch and Jan-\r{A}ke Larsson for useful discussions. D.R.M.A.-S. acknowledges support from the EPSRC, the Sweden-America Foundation, Hitachi Ltd, the Lars Hierta Memorial Foundation and Girton College. N.Y.H. was supported by an NSF grant for the Institute for Theoretical Atomic, Molecular, and Optical Physics at Harvard University and the Smithsonian Astrophysical Observatory. H.V.L. received funding from the European Union's Horizon 2020 research and innovation programme under the Marie Sk\l{}odowska-Curie grant agreement No 642688. A.A.L. acknowledges support from the EPSRC and Hitachi Ltd. S.L. was supported by NSF, AFOSR, and by ARO under the Blue Sky Initiative.

\bibliography{PSDraft}

\onecolumngrid

\newpage
\clearpage

    \section*{Supplementary Note 1 -- Expressing the postselected quantum Fisher information in terms of the KD distribution} \label{App:QuasDer}
\color{black}As shown in the Results section of our main paper, the postselected quantum Fisher information is given by
\begin{equation}
\label{Eq:FishQpsShort}
\mathcal{I}_Q(\theta|\Psi_{\theta}^\ps) =   4\braket{\dot{\psi}_{\theta}^\ps|\dot{\psi}_{\theta}^\ps} \frac{1}{p_{\theta}^\ps} -  4 |\braket{\dot{\psi}_{\theta}^\ps|\psi_{\theta}^\ps}|^2 \frac{1}{(p_{\theta}^\ps)^2} ,
\end{equation}
where nonrenormalized postselected quantum state is $\ket{\psi_{\theta}^\ps} = \hat{F} \hat{U}(\theta) \ket{\Psi_0} $, where $\ket{\Psi_0}\bra{\Psi_0} \equiv \hat{\rho}_0$. $p_{\theta}^\ps = \mathrm{Tr}(\hat{F}\hat{\rho}_{\theta})$ is the probability of postselection. 

In this supplementary note, we show that Eq. \ref{Eq:FishQpsShort} can be expressed in terms of the double-extended KD distribution:
\begin{align}
\label{Eq:FishQuas}
\mathcal{I}_Q(\theta | \Psi_{\theta}^\ps) =  4  \sum_{
\substack{a,a^{\prime}, \\ f\in \mathcal{F}^\ps } 
} \frac{q_{a,a^{\prime},f}^{\hat{\rho}_{\theta}}}{p_{\theta}^{\ps}}   a a^{\prime}   - 4   \Big| \sum_{
\substack{a,a^{\prime}, \\ f\in \mathcal{F}^\ps } 
} \frac{q_{a,a^{\prime},f}^{\hat{\rho}_{\theta}}}{p_{\theta}^{\ps}}  a \Big|^2 ,
\end{align} \color{black}

 The first term of the quantum Fisher information (Eq. \ref{Eq:FishQpsShort}) is 
\begin{align}
 \frac{4}{p_{\theta}^{\ps}} \braket{\dot{\psi}_{\theta}^{\ps} | \dot{\psi}_{\theta}^{\ps}}
& =  \frac{4}{p_{\theta}^{\ps}} \mathrm{Tr}\big( \hat{F}\dot{\hat{U}}(\theta)  \hat{\rho}_0 \dot{\hat{U}}^{\dagger}(\theta)  \hat{F}^{\dagger} \big) 
=  \frac{4}{p_{\theta}^{\ps}} \mathrm{Tr}\Big( \hat{F} \hat{A} \hat{\rho}_{\theta} \hat{A} \Big) \label{Eq:AppTr1}
 \\ 
& =  \frac{4}{p_{\theta}^{\ps}} \mathrm{Tr}\Big(  \sum_a{\ket{a}\bra{a}} a \hat{\rho}_{\theta}  
\sum_{a^{\prime}}{\ket{a^{\prime}}\bra{a^{\prime}}}a^{\prime}  \sum_{f\in \mathcal{F}_{\ps}}{\ket{f}\bra{f}}  \Big) , \label{Eq:QuasEigenReso}
\end{align}
where, in Eq. \ref{Eq:QuasEigenReso}, we have expressed $\hat{A}$ and $\hat{F}$ in their corresponding eigendecompositions. This expression can be rewritten in terms of the doubly extended Kirkwood-Dirac quasiprobability distribution \color{black} ($q_{a,a^{\prime},f}^{\hat{\rho}} = \braket{f|a}\bra{a}\hat{\rho} \ket{a^{\prime}} \braket{a^{\prime} | f} $)\color{black}:
\begin{align}
\frac{4}{p_{\theta}^{\ps}} \sum_{
\substack{a,a^{\prime}, \\ f\in \mathcal{F}^\ps } 
} \mathrm{Tr}\Bigg(   a a^{\prime} q_{a,a^{\prime},f}^{\hat{\rho}_{\theta}} \frac{\ket{a}\bra{f}}{\braket{f|a}} \Bigg)
=  \frac{4}{p_{\theta}^{\ps}}  \sum_{
\substack{a,a^{\prime}, \\ f\in \mathcal{F}^\ps } 
} q_{a,a^{\prime},f}^{\hat{\rho}_{\theta}}  a a^{\prime} .
\end{align}

Similarly, the second term of Eq. \ref{Eq:FishQpsShort} is
\begin{align}
 \frac{4}{(p_{\theta}^{\ps})^2} \big| \braket{\psi_{\theta}^{\ps} | \dot{\psi}_{\theta}^{\ps}} \big| ^2
=  \frac{4}{(p_{\theta}^{\ps})^2} \big|  \mathrm{Tr}\big(  \hat{F} \hat{\rho}_{\theta} \hat{A} \big)\big|^2
=  \frac{4}{(p_{\theta}^{\ps})^2}   \Big| \sum_{
\substack{a,a^{\prime}, \\ f\in \mathcal{F}^\ps } 
} q_{a,a^{\prime},f}^{\hat{\rho}_{\theta}}  a \Big|^2 . \label{Eq:AppTr2}
\end{align} 

Combining the expressions above gives Eq. \ref{Eq:FishQuas}:
\begin{align}
\mathcal{I}_Q(\theta | \Psi_{\theta}^\ps) =  4  \sum_{
\substack{a,a^{\prime}, \\ f\in \mathcal{F}^\ps } 
} \frac{q_{a,a^{\prime},f}^{\hat{\rho}_{\theta}}}{p_{\theta}^{\ps}}   a a^{\prime}   - 4   \Big| \sum_{
\substack{a,a^{\prime}, \\ f\in \mathcal{F}^\ps } 
} \frac{q_{a,a^{\prime},f}^{\hat{\rho}_{\theta}}}{p_{\theta}^{\ps}}  a \Big|^2 .
\end{align}

\section*{Supplementary note 2 -- Proof of Theorem 2} \label{App:PSFishBound}

Here, we prove Theorem 2. First, we upper-bound the right-hand side of Eq. \ref{Eq:FishQuas}, assuming that all $q_{a,a^{\prime},f}^{\hat{\rho}_{\theta}} / p_{\theta}^{\ps} \in [0, \, 1] $. We label the $M$ eigenvalues of $\hat{A}$ and arrange them in increasing order: $a_1,a_2,...,a_M$, such that $a_1 \equiv a_{\textrm{min}}$ and $a_M \equiv a_{\textrm{max}}$. Initially, we assume that the $0$-point of the eigenvalue axis is set such that $a_1 = 0$ and $a_M = \Delta a$. In this scenario, all the components of the first term of Eq. \ref{Eq:FishQuas} are nonnegative. We temporarily ignore the form of $q_{a,a^{\prime},f}^{\hat{\rho}_{\theta}} / p_{\theta}^{\ps}$, and treat this ratio as a general quasiprobability distribution. Then, $\mathcal{I}_Q(\theta | \Psi_{\theta}^\ps)$ maximizes when $ q_{a,a^{\prime},f}^{\hat{\rho}_{\theta}} / p_{\theta}^{\ps} $ vanishes at all $a^{\prime}$ values except  $a^{\prime} = a_{\textrm{max}}$. We define $q_a \equiv \sum_{a^{\prime}, f\in \mathcal{F}^\ps } q_{a,a^{\prime},f}^{\hat{\rho}_{\theta}} / p_{\theta}^{\ps} $, such that  all $q_a  \in [0, \, 1] $ and $\sum_{a } q_a = 1 $. If $ q_{a,a^{\prime},f}^{\hat{\rho}_{\theta}} / p_{\theta}^{\ps} $ is nonzero only when $a^{\prime} = a_{\textrm{max}}$, Eq. \ref{Eq:FishQuas} becomes
\begin{align}
\mathcal{I}_Q(\theta | \Psi_{\theta}^\ps) =  4  a_{M} \sum_{a} q_a a    - 4   \Big( \sum_{a}q_a  a \Big)^2 .
\end{align}
Expanding each sum, we obtain
\begin{align}
\mathcal{I}_Q(\theta | \Psi_{\theta}^\ps)  & =  4  a_{M} (q_{a_1} a_1 + K + q_{a_M} a_M ) - 4   (q_{a_1} a_1 + K + q_{a_M}  a_M )^2 \\  & =  4  a_{M} ( K + q_{a_M} a_M )   - 4   ( K + q_{a_M}  a_M )^2 ,\label{Eq:NonNegFish2}
\end{align}
where we used $q_{a_1} a_1 =  0$ and defined $K \equiv \sum_{a\in \{ a_2,...,a_{M-1} \} } q_a a \leq a_M$. As $\hat{A}$ is not totally degenerate, $a_M \neq 0$, and Eq. \ref{Eq:NonNegFish2} is maximized when $ q_{a_M} = ( a_M - 2K ) / (2 a_M) $. This yields
\begin{align}
\max \{ \mathcal{I}_Q(\theta | \Psi_{\theta}^\ps) \} =  a_M^2 = (\Delta a)^2 ,
\end{align}
where we have recalled that $a_M  = \Delta a$. 

We are left with proving that we can always set $a_1 = 0$ and $a_M  = \Delta a$. We continue to assume that $q_{a,a^{\prime},f}^{\hat{\rho}_{\theta}} / p_{\theta}^{\ps} \in [0, \, 1] $, and we shift all the eigenvalues by a constant real value $ \delta_a$. The effect on $\mathcal{I}_Q(\theta | \Psi_{\theta}^\ps)$ is
\begin{gather}
\mathcal{I}_Q(\theta | \Psi_{\theta}^\ps) 
 \rightarrow  
4  \sum_{
\substack{a,a^{\prime}, \\ f\in \mathcal{F}^\ps } 
} \frac{q_{a,a^{\prime},f}^{\hat{\rho}_{\theta}}}{p_{\theta}^{\ps}}   (a+ \delta_a) (a^{\prime}+ \delta_a)   - 4   \Big[ \sum_{
\substack{a,a^{\prime}, \\ f\in \mathcal{F}^\ps } 
} \frac{q_{a,a^{\prime},f}^{\hat{\rho}_{\theta}}}{p_{\theta}^{\ps}}  (a+ \delta_a) \Big]^2 \\
=
 4  \sum_{
\substack{a,a^{\prime}, \\ f\in \mathcal{F}^\ps } 
} \frac{q_{a,a^{\prime},f}^{\hat{\rho}_{\theta}}}{p_{\theta}^{\ps}}   a a^{\prime}   - 4   \Big[ \sum_{
\substack{a,a^{\prime}, \\ f\in \mathcal{F}^\ps } 
} \frac{q_{a,a^{\prime},f}^{\hat{\rho}_{\theta}}}{p_{\theta}^{\ps}}  a \Big]^2 +  4 \delta_a \Big( \sum_{
\substack{a,a^{\prime}, \\ f\in \mathcal{F}^\ps } 
} \frac{q_{a,a^{\prime},f}^{\hat{\rho}_{\theta}}}{p_{\theta}^{\ps}}   a -\sum_{
\substack{a,a^{\prime}, \\ f\in \mathcal{F}^\ps } 
} \frac{q_{a,a^{\prime},f}^{\hat{\rho}_{\theta}}}{p_{\theta}^{\ps}}   a^{\prime} \Big)  = \mathcal{I}_Q(\theta | \Psi_{\theta}^\ps) . \label{Eq:ShiftFish}
\end{gather}
The last equality holds because $q_{a,a^{\prime},f}^{\hat{\rho}}  = \big( q_{a^{\prime},a,f}^{\hat{\rho}} \big)^{*}$ generally and we are assuming that $q_{a,a^{\prime},f}^{\hat{\rho}} \in \mathbb{R}$. Consequently, if all $q_{a,a^{\prime},f}^{\hat{\rho}_{\theta}} / p_{\theta}^{\ps} \in [0, \, 1] $, then $\mathcal{I}_Q(\theta | \Psi_{\theta}^\ps) \leq (\Delta a)^2 $. \color{black} The second term of Eq. \ref{Eq:FishQuas} cannot be decreased by imaginary values in $q_{a,a^{\prime},f}^{\hat{\rho}_{\theta}}$. Moreover, the first term is necessarily real and nonnegative. Thus imaginary elements $q_{a,a^{\prime},f}^{\hat{\rho}_{\theta}}$ cannot increase $\mathcal{I}_Q(\theta | \Psi_{\theta}^\ps)$. \color{black}  If $\mathcal{I}_Q(\theta | \Psi_{\theta}^\ps) > (\Delta a)^2 $, then $q_{a,a^{\prime},f}^{\hat{\rho}_{\theta}}$ must have negative  entries. $\square$
    
      \section*{Supplementary note 3 -- Infinite postselected quantum Fisher information} \label{App:InfPSFish}
 
Here, we show that the postselected quantum Fisher information $\mathcal{I}_Q(\theta | \Psi_{\theta}^\ps)$ can approach infinity. The proof is by example; other examples might exist.

We assume that the generator $\hat{A}$ has $M \geq 3$ eigenvalues that are not all identical. We also assume that we possess an estimate $\theta_0$ that lies close to the true value of $\theta$: $ \delta_{\theta} \equiv \theta - \theta_0 $, with $|\delta_{\theta}| \ll 1$. (The derivation of the quantum Fisher information also rests on the assumption that one has access to such an estimate \cite{Braunstein94}.)  

By Eqs. \ref{Eq:FishQpsShort}, \ref{Eq:AppTr1} and \ref{Eq:AppTr2},
\begin{equation}
\label{Eq:LongPSqFish}
\mathcal{I}_Q(\theta | \Psi_{\theta}^{\textrm{ps}}) = \frac{4}{p_{\theta}^{\ps}} \mathrm{Tr}\Big( \hat{F} \hat{A} \hat{U}(\theta) \hat{\rho}_{0} \hat{U}(\theta)^{\dagger} \hat{A} \Big) -  \frac{4}{(p_{\theta}^{\ps})^2} \Big|  \mathrm{Tr}\Big(  \hat{F} \hat{U}(\theta) \hat{\rho}_{0} \hat{U}(\theta)^{\dagger} \hat{A} \Big)\Big|^2 .
\end{equation}
We now choose $\hat{F}$ and $\hat{\rho}_{0}$ such that $\mathcal{I}_Q(\theta | \Psi_{\theta}^{\textrm{ps}})$ approaches infinity. Crudely, $p_{\theta}^{\ps}$ must approach $0$ while $\mathrm{Tr} ( \hat{F} \hat{A} \hat{U}(\theta) \hat{\rho}_{0} \hat{U}(\theta)^{\dagger} \hat{A} )$ either stays constant or approaches $0$ more slowly. We label the $M$ eigenvalues of $\hat{A}$ and arrange them in increasing order: $a_1,a_2,...,a_M$, such that $a_1 \equiv a_{\textrm{min}}$ and $a_M \equiv a_{\textrm{max}}$.


First, we choose $\hat{F} = \ket{f_1}\bra{f_1}+\ket{f_2}\bra{f_2}$, where 
\begin{align}
& \ket{f_1} \equiv \frac{\ket{a_{\max}} + \ket{a_{\min}}}{\sqrt{2}} , \\
& \ket{f_2} \equiv \frac{ \frac{i}{\sqrt{2}} (\ket{a_{\max}} - \ket{a_{\min}}) + \ket{a_k} }{\sqrt{2}} ,
\end{align}
and $\ket{a_k} \neq \ket{a_{\max}}, \; \ket{a_{\min}} $ . We also choose $\hat{\rho}_{0} = \ket{\Psi_0} \bra{\Psi_0}$ such that
\begin{align}
\ket{\Psi_0}  \equiv \ket{\Psi_0(\theta_0, \phi)} = \hat{U}^{\dagger}(\theta_0)\frac{1}{\sqrt{2}} \Big\{ & [\cos{(\phi)} - \sin{(\phi)}] \frac{i}{\sqrt{2}} (\ket{a_{\min}} - \ket{a_{\max}})   + [\cos{(\phi)} + \sin{(\phi)}] \ket{a_k} \Big\} .
\end{align}
 $\phi \approx 0$ is a parameter that can be tuned to maximize the postselected Fisher information for a given approximation accuracy $ \delta_{\theta}$. As $\phi$ is a parameter of the input state, variations in the Fisher information with $\phi$ will reflect the effects of disturbances to the input state. Substituting the expressions for $\hat{F}$ and $\hat{\rho}_{0}$ into Eq. \ref{Eq:LongPSqFish}, we find 
\begin{align}
\label{Eq:psOptimalFish}
\mathcal{I}_Q(\theta | \Psi_{\theta}^{\textrm{ps}}) = & \; 8 \big\{ 5 - 
   2 \cos(2 \phi) \big(\cos[(a_{M} - a_{k}) \delta_{\theta}] + 
      \cos[(a_{k} - a_{1}) \delta_{\theta}] \big) + 
   \cos[(a_{M} - a_{1}) \delta_{\theta}] [ \sin(2 \phi) - 1] - 
   \sin(2 \phi) \big\}^{-2}
   \nonumber \\
   &
   \times 
   \Big\{ 2 a_{M}^2 - 
     a_{M} a_{k} + a_{k}^2 + 
     2 a_{1}^2 - (3 a_{M} + a_{k}) a_ {1} + (a_{M} - a_{k}) (a_{k} - a_{1}) \cos(
       4 \phi) \big( \cos[(a_{M} - a_{1}) \delta_{\theta}] - 1 \big) 
       \nonumber \\
        & 
        + (a_{M} - 
        a_{k}) (a_{k} - a_{1}) \cos[(a_{M} - a_{1}) \delta_{\theta}] 
         +
     2  (a_{M} - a_{1}) \cos(
       2 \phi) \big( (a_ {1} - 
           a_{k}) \cos[(a_ {M} - a_{k}) \delta_{\theta}] 
           \nonumber \\
        & 
        + 
        (a_{k} - 
           a_{M}) \cos[(a_ {k} - a_{1}) \delta_{\theta}] \big) - 
     2 (a_{M} - a_{1})^2 \sin(
       2 \phi) + (a_{M} - 
        a_{1}) \big( (a_{k} - 
           a_{1}) \cos[(a_{M} - a_{k}) \delta_{\theta}]
           \nonumber \\
        & 
         + (a_{M} - 
           a_{k}) \cos[(a_{k} - a_{1}) \delta_{\theta}] \big) \sin(
       4 \phi) \Big\} .
\end{align}
The postselection probability is
\begin{align}
 p^{\textrm{ps}}_{\theta} =  \frac{1}{8} \Big\{ 5 - 
   2 \cos(2 \phi) \big(\cos[(a_{M} - a_{k}) \delta_{\theta}] + 
      \cos[(a_{k} - a_{1}) \delta_{\theta}] \big) + 
   \cos[(a_M - a_1) \delta_{\theta}] [ \sin(2 \phi) - 1] - \sin(2 \phi) \Big\} .
\end{align}

In the limit as our estimate $\theta_0$ approaches the true value of $\theta$, such that $\delta_{\theta} \rightarrow 0$, 
\begin{align}
\lim_{\delta_{\theta} \rightarrow 0}  p^{\textrm{ps}}_{\theta}  & = \sin^2(\phi) , \\
\lim_{\delta_{\theta} \rightarrow 0}  \mathcal{I}_Q(\theta | \Psi_{\theta}^{\textrm{ps}})  & = \frac{(\cot{(\phi)} - 1)^2}{2} (\Delta a)^2 , \; \; \textrm{and}\\
\lim_{\delta_{\theta} \rightarrow 0} p^{\textrm{ps}}_{\theta}  \times \mathcal{I}_Q(\theta | \Psi_{\theta}^{\textrm{ps}})  & = \frac{1}{2} [1-\sin{(2 \phi)}] (\Delta a)^2 .
\end{align}
In the limit as $\phi \rightarrow 0$,
\begin{align}
\lim_{  \phi \rightarrow 0 } \Big[\lim_{\delta_{\theta} \rightarrow 0}  p^{\textrm{ps}}_{\theta} \Big] & = 0  , \\
\lim_{  \phi \rightarrow 0  } \Big[ \lim_{\delta_{\theta} \rightarrow 0}  \mathcal{I}_Q(\theta | \Psi_{\theta}^{\textrm{ps}}) \Big] & = \infty  , \; \; \textrm{and} \label{Eq:ArbFishPS} 
 \\
\lim_{  \phi \rightarrow 0 } \Big[ \lim_{\delta_{\theta} \rightarrow 0} p^{\textrm{ps}}_{\theta}  \times \mathcal{I}_Q(\theta | \Psi_{\theta}^{\textrm{ps}}) \Big] & = \frac{1}{2} (\Delta a)^2 . 
\end{align}
According to Eq. \ref{Eq:ArbFishPS}, if first $\delta_{\theta}$ and then $\phi$ approaches $0$ in Eq. \ref{Eq:psOptimalFish}, $\mathcal{I}_Q(\theta | \Psi_{\theta}^{\textrm{ps}})$ approaches infinity. 

\begin{figure}
\includegraphics[scale=0.4]{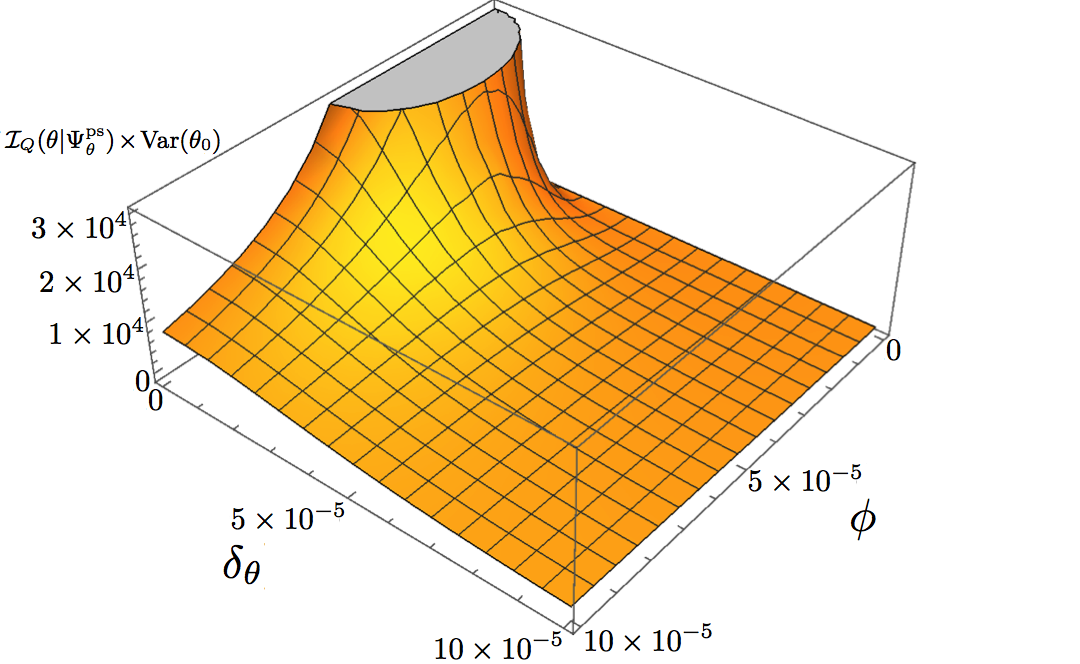}
\caption{\color{black} \textbf{Scaled postselected quantum Fisher information.} The figure shows the   postselected quantum Fisher information (Eq. \ref{Eq:psOptimalFish}) multiplied by the pre-experiment variance $\textrm{Var}( \theta_0 ) $ as a function of $\phi$ and $\delta_{\theta}$. \color{black}  For small values of $\delta_{\theta}$ and $\phi$, the value of $\mathcal{I}_Q(\theta | \Psi_{\theta}^{\textrm{ps}}) \times \textrm{Var}( \theta_0 )$ diverges. The eigenvalues $a_1$, $a_k$ and $a_M$ are set to $-1$, $1$ and $3$, respectively. $\textrm{Var}( \theta_0 )$ was set to $1 \times 10^{-6}$.}
\label{fig:OptimalFishEst}
\end{figure}

There are a few points to note. First, $\mathcal{I}_Q(\theta | \Psi_{\theta}^{\textrm{ps}})$ diverges in the two ordered limits. In any real experiment, one could not blindly set $\phi = 0$, but would have to choose $\phi$ based on an estimate of $\theta$. Second, if $\delta_{\theta} \approx 0$, then $\theta_0 \approx \theta$, and the pre-experiment variance of our initial estimate $\theta_0$, $\textrm{Var}( \theta_0 )$, must be small. That is, we begin the experiment with much information about $\theta$. Guided by the Cramér-Rao bound, we expect that, in a useful experiment,  $\mathcal{I}_Q(\theta | \Psi_{\theta}^{\textrm{ps}})$ would grow large, while $1 / \textrm{Var}( \theta_0 ) < \mathcal{I}_Q(\theta | \Psi_{\theta}^{\textrm{ps}})$.   Figure \ref{fig:OptimalFishEst} shows $\mathcal{I}_Q(\theta | \Psi_{\theta}^{\textrm{ps}}) \times \textrm{Var}( \theta_0 )$  as a function of $\phi$ and $\delta_{\theta}$ for an experiment where $a_1 = -1 $, $a_k = 1 $, $a_M = 3 $ and $\textrm{Var}( \theta_0 ) =  10^{-6}$. If $\theta_0$ is within a few $\sigma_{\theta_0} \equiv \sqrt{\textrm{Var}( \theta_0 )}$ of $\theta$, then $\mathcal{I}_Q(\theta | \Psi_{\theta}^{\textrm{ps}}) \times \textrm{Var}( \theta_0 ) \gg 1$.  Figure \ref{fig:OptimalFishEst} shows that large values of $1 / \delta_{\theta}$ can result in even larger values of $\mathcal{I}_Q(\theta | \Psi_{\theta}^{\textrm{ps}})$.  Figure \ref{fig:OptimalFishEst} also illustrates the effect of input-state disturbances of $\phi$ on $\mathcal{I}_Q(\theta | \Psi_{\theta}^{\textrm{ps}}) \times \textrm{Var}( \theta_0 )$. Third, while the theoretical strategy investigated in this appendix achieves an infinite postselected quantum Fisher information, the postselection also ``wastes'' information as $\lim_{  \phi \rightarrow 0 } [ \lim_{\delta_{\theta} \rightarrow 0} p^{\textrm{ps}}_{\theta}  \times \mathcal{I}_Q(\theta | \Psi_{\theta}^{\textrm{ps}}) ]< (\Delta a)^2 $. If $\hat{A}$ possesses certain properties, it is possible to avoid wasting information through the postselection; we show how in the following appendix.

      \section*{Supplementary note 4 --  Infinite postselected quantum Fisher information without loss of information} \label{App:InfOptPSFish}

If the generator $\hat{A}$ has $M \geq 4$ eigenvalues, and the minimum and maximum eigenvalues are both at least doubly degenerate, then $\mathcal{I}_Q(\theta | \Psi_{\theta}^{\textrm{ps}})$ can approach infinity without information's being lost in the events discarded by postselection. We show how below. 

First, we assign the orthonormal eigenvectors $\ket{a_{\min_1}}$ and $\ket{a_{\min_2}}$ to the eigenvalues $a_1 = a_{\min}$ and $a_2 = a_{\min}$, respectively. Here, we have reused the eigenvalue notation from Supp. Mat. \ref{App:InfPSFish}. Similarly, we assign the orthonormal eigenvectors $\ket{a_{\max_1}}$ and $\ket{a_{\max_2}}$ to the eigenvalues $a_M = a_{\max}$ and $a_{M-1} = a_{\max}$, respectively. Second, we set $\hat{F} = \ket{f_1}\bra{f_1}+\ket{f_2}\bra{f_2}$, where 
\begin{align}
& \ket{f_1} \equiv \frac{ \ket{ a_{\max_2} } - \ket{a_{\min_1}} }{\sqrt{2}} , \\
& \ket{f_2} \equiv \frac{ \ket{a_{\min_2}} - \ket{a_{\max_1}} }{\sqrt{2}} .
\end{align}
We also choose $\ket{\Psi_0}$ such that
\begin{align}
\ket{\Psi_0 (\theta_0, \phi)} =  \hat{U}^{\dagger}(\theta_0) \frac{1}{2} \big\{ [ \cos{(\phi)} - \sin{(\phi)} ]  ( \ket{ a_{\max_2} } + \ket{ a_{\min_2} } )  + [ \sin{(\phi)} + \cos{(\phi)} ]  ( \ket{ a_{\max_1} } + \ket{ a_{\min_1} } ) \big\} .
\end{align}
As in App. \ref{App:InfPSFish}, $\phi \approx 0$ is a parameter that can be tuned to maximize $\mathcal{I}_Q(\theta | \Psi_{\theta}^{\textrm{ps}})$ for a given approximation accuracy of $ \delta_{\theta}$.

Substituting the expressions for $\hat{F}$ and $\hat{\rho}_{0}$ into Eq. \ref{Eq:LongPSqFish}, we find 
\begin{align}
\label{Eq:psOptimalFishOptimalScale}
\mathcal{I}_Q(\theta | \Psi_{\theta}^{\textrm{ps}}) = \frac{\sin^2{(2 \phi)} (a_M - a_1 )^2 }{\big(  1 - \cos{(2 \phi)} \cos{[(a_M - a_1) \delta_{\theta}]}          \big)^2} .
\end{align}
The postselection probability is
\begin{align}
 p^{\textrm{ps}}_{\theta} =  \frac{1}{2} \Big\{ 1 - 
    \cos(2 \phi) \cos[(a_{M} - a_{1}) \delta_{\theta}] \Big\} .
\end{align}

Again, we  investigate the limit as our estimate $\theta_0$ approaches the true value of $\theta$:
\begin{align}
\lim_{\delta_{\theta} \rightarrow 0}  p^{\textrm{ps}}_{\theta}  & = \sin^2(\phi) , \\
\lim_{\delta_{\theta} \rightarrow 0}  \mathcal{I}_Q(\theta | \Psi_{\theta}^{\textrm{ps}})  & = \cot^2{(\phi)} (\Delta a)^2 ,    \; \; \textrm{and}  \\
\lim_{\delta_{\theta} \rightarrow 0} p^{\textrm{ps}}_{\theta}  \times \mathcal{I}_Q(\theta | \Psi_{\theta}^{\textrm{ps}})  & = \cos^2{(\phi)} (\Delta a)^2 .
\end{align}
In the limit as $\phi \rightarrow 0$,
\begin{align}
\lim_{  \phi \rightarrow 0 } \Big[\lim_{\delta_{\theta} \rightarrow 0}  p^{\textrm{ps}}_{\theta} \Big] & = 0  , \\
\lim_{  \phi \rightarrow 0  } \Big[ \lim_{\delta_{\theta} \rightarrow 0}  \mathcal{I}_Q(\theta | \Psi_{\theta}^{\textrm{ps}}) \Big] & = \infty    , \; \; \textrm{and}  
 \\
\lim_{  \phi \rightarrow 0 } \Big[ \lim_{\delta_{\theta} \rightarrow 0} p^{\textrm{ps}}_{\theta}  \times \mathcal{I}_Q(\theta | \Psi_{\theta}^{\textrm{ps}}) \Big] & =  (\Delta a)^2 . 
\end{align}
In conclusion, the above strategy allows us to obtain an infinite value for $\mathcal{I}_Q(\theta | \Psi_{\theta}^{\textrm{ps}})$, while $p^{\textrm{ps}}_{\theta}  \times \mathcal{I}_Q(\theta | \Psi_{\theta}^{\textrm{ps}}) = (\Delta a)^2 $. No information is lost in the postselection. As in Supp Mat. \ref{App:InfPSFish}, the results hold for the two ordered limits.

\end{document}